\documentclass[12pt]{article}

\usepackage{amssymb}
\usepackage{amsmath}
\usepackage{latexsym}
\usepackage{yfonts}

\catcode`@=11
\renewcommand{\section}{\@startsection{section}{1}{0pt}{\medskipamount}
{\medskipamount}{\large\bf}} \numberwithin{equation}{section}
\catcode`@=12

\def\beq{\begin{eqnarray}}    
\def\eeq{\end{eqnarray}}      

\def\pa{\partial}                       

\def\={\ =\ }

\begin{document}

\begin{center}

{\Large\bf Covariant quantization of Yang-Mills theory in the first order formalism}

\vspace{18mm}

{\large
P.M. Lavrov$^{(a, b)}\footnote{E-mail:
lavrov@tspu.edu.ru}$\; }

\vspace{8mm}

\noindent  ${{}^{(a)}} ${\em
Tomsk State Pedagogical University,\\
Kievskaya St.\ 60, 634061 Tomsk, Russia}

\noindent  ${{}^{(b)}} ${\em
National Research Tomsk State  University,\\
Lenin Av.\ 36, 634050 Tomsk, Russia}

\vspace{20mm}

\begin{abstract}
\noindent
In the present paper the Yang-Mills theory in the first order formalism
is studied. On classical level the first order formulation is equivalent
to the standard second order description of the Yang-Mills theory.
It is proven that both formulations remain equivalent on quantum level as well.

\end{abstract}

\end{center}

\vfill

\\

\newpage

\section{Introduction}
\noindent
Yang-Mills theories have always attracted
(see, for example,  the famous textbook by Weinberg \cite{Weinberg} for qualitative
discussions and presentations of numerous aspects
of classical and quantum properties of Yang-Mills fields)
and continue to attract
\cite{Barv,BLT'18,BLT'19,BFMc2} the attention of researchers, since they play
a key role in the mainstream
of modern models of fundamental interactions. In contrast with the electrodynamics
the Yang-Mills theories belong to non-abelian gauge theories. It caused the problem
indicated firstly by Feynman \cite{Feynman} and related to the S-matrix unitarity
in Yang-Mills theories when simple modification of quantization rules adopted in
quantum electrodynamics is used. Later on the correct quantization of Yang-Mills
theories has been found \cite{FP,DeWitt1}.

Recently, there has been activity in the study of quantum properties of the
Yang-Mills theory formulated in the first order formalism
\cite{BMc,BFMc1,McFM-FB,McBFM-F}
instead of standard approach based on the second order form \cite{FP}.
On classical level both formulations are equivalent. It was really assumed in
\cite{McFM-FB,McBFM-F} that all Green functions appeared in the standard
quantization can be reproduced with the help of using more simple first order formulation.
This proposal is closely related to quantum equivalence of both presentations of the
Yang-Mills theory. In our knowledge the quantum equivalence of the Yang-Mills theory
presented in the first and second orders was not studied before in
the scientific literature.

In the present paper we study the Yang-Mills theory in the first order formulation.
This  formulation requires introduction additional
antisymmetric second order tensor
fields, ${\cal F}^a_{\mu\nu}$. In turn it leads to appearing additional
gauge symmetry.  Further quantization depends on properties of this additional gauge
symmetry and on the gauge algebra to be closed/open and irreducible/reducible.
Then the found structure of gauge algebra allows to construct a quantum action using,
in general, the BV-formalism \cite{BV,BV1}. Fixing a gauge via the standard BV procedure
one obtains the full quantum action which is used
to define the generating functional of Green functions in
the form of functional integral. Quantum equivalence  or non-equivalence of
the Yang-Mills theory in two formulations
can be studied by comparison  of corresponding vacuum functionals. If the vacuum
functionals coincide then the quantum equivalence is realized. Otherwise
one meets the quantum non-equivalence.

The paper is organized as follows. In Sec. 2 gauge symmetries of the Yang-Mills
theory written in the first order formulation are studied.
In Sec. 3 the full quantum action found as solution to the classical master equation
of the BV-formalism with applying the standard gauge fixing procedure in linear
Lorentz invariant gauges and the BRST symmetry are found.
In Sec. 4 the generating functional of Green functions in the
form of functional integral and the Ward identities  are constructed.
In Sec. 5 the quantum equivalence of two approaches for quantization of the Yang-Mills theory
is proven.
Finally, in Sec. 6 the results obtained in the paper are discussed.

In the paper the DeWitt's condensed notations are used \cite{DeWitt}.
The right and left functional derivatives with respect to
fields and antifields are marked by special symbols $"\leftarrow"$  and
$"\rightarrow"$ respectively. Arguments of any functional are enclosed
in square brackets
$[\;]$, and arguments of any function are enclosed in parentheses, $(\;)$.

\setcounter{section}{1}
\renewcommand{\theequation}{\thesection.\arabic{equation}}
\setcounter{equation}{0}

\section{ Gauge symmetries}
\noindent
Standard formulation of the Yang-Mills fields, $A^a_{\mu}$,
operates with the second order action
\beq
\label{YM2}
S^{(2)}[A] &=&-\frac{1}{4}
F_{\mu \nu }^{a}(A)F^{a\mu \nu}(A),   \label{YM}
\eeq
where $F_{\mu \nu }^{a}(A)$ is the field strength
\beq
F_{\mu \nu }^{a }(A) &=&\partial _{\mu }A_{\nu }^{a }-\partial
_{\nu }A_{\mu }^{\alpha }+f^{abc }A_{\mu }^{b}A_{\nu }^{c},
\eeq
and $f^{abc }$ are completely antisymmetric structure constants of the Lie algebra
satisfying the Jacobi identity
\beq
f^{abc}f^{cde}+f^{aec}f^{cbd}+f^{adc}f^{ced}\equiv 0.
\eeq
The equations of motion read
\beq
\label{EOM2}
\frac{\delta S^{(2)}[A]}{\delta A^a_{\mu}}=D^{ab}_{\nu}(A)F^{b\nu\mu}(A)=0,
\eeq
where  $D^{ab}_{\mu}(A)$ is the covariant derivative
\beq
D^{ab}_{\mu}(A)=\delta^{ab}\pa_{\mu}+f^{acb}A^c_{\mu}.
\eeq

The action (\ref{YM2}) is invariant under the gauge transformations of $A^a_{\mu}$,
\beq
\label{gt2}
\delta_{\xi}S^{(2)}[A]=0,\qquad \delta_{\xi} A^a_{\mu}=D^{ab}_{\mu}(A)\xi^b,
\eeq
where  $\xi^a$ are arbitrary functions of space-time coordinates. Notice that
under the gauge transformations the field strength
tensor transforms by the tensor law,
\beq
\delta_{\xi} F_{\mu \nu }^{a}(A)=f^{abc}F_{\mu \nu }^{b}(A)\xi^c.
\eeq
Algebra of gauge transformations
\beq
\label{ga2s}
[\delta_{\xi_1}, \delta_{\xi_2}]A^a_{\mu}=D^{ab}_{\mu}(A)\xi^b_3,\quad
\xi^a_3=f^{abc}{\xi}^b_1{\xi}^c_2,
\eeq
is closed and irreducible.

The first order formulation of Yang-Mills fields is based on the action
\beq
\label{YM1}
S^{(1)}[A, {\cal F}] =\frac{1}{4}
{\cal F}_{\mu \nu }^ {a}{\cal F}^{a\mu \nu }-\frac{1}{2}
F_{\mu \nu }^{a}(A){\cal F}^{a\mu \nu },
\eeq
where ${\cal F}_{\mu \nu }^ {a}$ are new antisymmetric tensor fields,
${\cal F}_{\mu \nu }^ {a}=-{\cal F}_{\nu \mu }^ {a}$.
The equations of motion read
\beq
\label{EOM1}
\frac{\delta S^{(1)}[A, {\cal F}]}{\delta A^a_{\mu}}=
D^{ab}_{\nu}(A){\cal F}^{b\nu \mu }=0,\qquad
\frac{\delta S^{(1)}[A, {\cal F}]}{\delta {\cal F}_{\mu \nu }^{a}}=
\frac{1}{2}\big({\cal F}^{a\mu \nu }-F^{a\mu \nu }(A)\big)=0.
\eeq
From the second in (\ref{EOM1}) it follows
${\cal F}_{\mu \nu }^{a}=F_{\mu \nu }^{a}(A)$. Substituting this result
in the first of (\ref{EOM1}) one obtains (\ref{EOM2}). On classical level we have
two equivalent descriptions of Yang-Mills fields.

The action (\ref{YM1}) is invariant under the following gauge transformations
\beq
\label{gt1}
\delta_{\xi}S^{(1)}[A, {\cal F}] =0,\qquad
\delta_{\xi} A^a_{\mu}=D^{ab}_{\mu}(A)\xi^b,\quad
\delta_{\xi} {\cal F}_{\mu \nu }^{a}= f^{abc}{\cal F}_{\mu \nu }^{b}\xi^c,
\eeq
which are considered as generalization of (\ref{gt2}). Introduction of new fields may lead to
increasing the degrees of freedom and as consequence
to existence of additional gauge symmetry. Indeed, the action (\ref{YM1})
is invariant under the
additional gauge transformations
\beq
\label{gt1ad}
\delta_{{\bar \xi}}S^{(1)}[A, {\cal F}]=0,\qquad
\delta_{{\bar \xi}} A^a_{\mu}=0,\quad
\delta_{{\bar \xi}}{\cal F}_{\mu \nu }^{a}=
f^{abc}\big({\cal F}_{\mu \nu }^{b}-F_{\mu \nu }^{b}(A)\big){\bar \xi}^c,
\eeq
where ${\bar \xi}^a$ are arbitrary functions of space-time coordinates.
Due to the second in (\ref{EOM1}) these gauge transformations belong to the class
of trivial gauge transformations \cite{BV,BV1,Hen} which are not related
with real degeneracy of initial action.
Therefore the first order formulation of Yang-Mills fields is accompanying by the following
gauge algebra
\beq
\label{gal1}
[\delta_{\xi_1}, \delta_{\xi_2}]A^a_{\mu}=D^{ab}_{\mu}(A)\xi^b_3,\quad
[\delta_{\xi_1}, \delta_{\xi_2}]{\cal F}_{\mu \nu }^{a}=
f^{acb}{\cal F}_{\mu \nu }^{c}{\xi}^b_3,
\eeq
where $\xi^a_3$ is defined in (\ref{ga2s}).
The gauge algebra (\ref{gal1}) is closed and irreducible.

\section{Quantum action}
\noindent
In construction of quantum action for Yang-Mills theory in the first order form
we will follow prescriptions of the BV-formalism \cite{BV}. Minimal antisymplectic space
is defined by the set of fields $\phi^A_{min}$ and antifields $\phi^*_{A\;min}$
\beq
\phi^A_{min}=\big(A^{a\mu}, {\cal F}^{a\mu\nu},C^a\big),\quad
\phi^*_{A\;min}=\big(A^*_{a\mu}, {\cal F}^*_{a\mu\nu},C^*_a\big),
\eeq
where $C^a$  are the ghost fields corresponding to the gauge symmetries
(\ref{gt1}).  Due to closedness  and
irreducibility of gauge algebra the minimal quantum action,
$S_{min}=S[\phi_{min},\phi^*_{min}]$, being a solution to the classical
master equation\footnote{For any set of fields $\phi^A$ and antifields $\phi^*_A$
and any functionals $F$ and $G$ the antibracket is defined by the rule
$(F,G)=F\Big(\overleftarrow{\pa}_{\phi^A}\overrightarrow{\pa}_{\phi^*_A}-
\overleftarrow{\pa}_{\phi^*_A}\overrightarrow{\pa}_{\phi^A}\Big)G$. }
\beq
(S_{min},S_{min})=0,
\eeq
satisfying the boundary condition
\beq
S_{min}\big|_{\phi^*_{min}=0}=S^{(1)}[A,{\cal F}]
\eeq
can be found in the form linear in antifields as
\beq
S_{min}=S^{(1)}[A,{\cal F}]+A^*_{a\mu}D^{ab\mu}(A)C^b+
{\cal F}^*_{a\mu\nu}f^{abc}{\cal F}^{b\mu\nu}C^c-C^*_a\frac{1}{2}f^{abc}C^bC^c.
\eeq

Now the extended action $S=S[\phi^{(1)},\phi^{*(1)}]$ in full antisymplectic space
\beq
\phi^{A(1)}=\big(A^{a\mu}, {\cal F}^{a\mu\nu},C^a,{\bar C}^a,
B^a\big),\quad
\phi^{*(1)}_{A}=\big(A^*_{a\mu}, {\cal F}^*_{a\mu\nu},C^*_a,
{\bar C}^*_a,,
B^*_a\big),
\eeq
has the form
\beq
\nonumber
S=S^{(1)}[A,{\cal F}]+A^*_{a\mu}D^{ab\mu}(A)C^b+
{\cal F}^*_{a\mu\nu}f^{abc}{\cal F}^{b\mu\nu}C^c
-\frac{1}{2}C^*_a f^{abc}C^bC^c+
{\bar C}^*_aB^a
\eeq
and satisfies the classical master equation and the boundary condition
\beq
(S,S)=0, \quad S\big|_{\phi^{*(1)}=0}=S^{(1)}[A,{\cal F}].
\eeq
Here ${\bar C}^a$  are the antighost fields to $C^a$
and $B^a$ are auxiliary fields (Nakanishi-Lautrup fields).
The gauge fixed action $S_{eff}[\phi]$ in the BV-formalism is defined as
\beq
S_{eff}[\phi^{(1)}]=S\big[\phi^{(1)},\phi^{*(1)}=
\Psi\overleftarrow{\pa}_{\phi^{(1)}}\big],
\eeq
where $\Psi=\Psi[\phi^{(1)}]$ is an odd gauge fixing functional.
For the closed and irreducible gauge algebra $S_{eff}[\phi^{(1)}]$ is usually refereed as
the Faddeev -Popov action $S_{FP}[\phi^{(1)}]$.
For the model under consideration the functional
$\Psi$ corresponding to the case of non-singular Lorentz
invariant and linear gauges can be chosen in the form
\beq
\Psi={\bar C}^a\chi^a(A,B),
\eeq
where $\chi^a(A,B)$  are gauge fixing functions,
\beq
\label{gff}
\chi^a(A,B)=\pa^{\mu}A^a_{\mu}+\frac{\xi}{2}B^a.
\eeq
In (\ref{gff})
$\xi$ is constant gauge parameters.

The Faddeev-Popov action for the model (\ref{YM1}) in the gauges (\ref{gff}) reads
\beq
\label{FP1}
S_{FP}^{(1)}[\phi^{(1)}]=S^{(1)}[A,{\cal F}]+{\bar C}^a\pa^{\mu}D^{ab}_{\mu}C^b+
B^a\chi^a(A,B).
\eeq

The Faddeev-Popov action for the model (\ref{YM2}) in gauges  (\ref{gff})
has the form
\beq
\label{FP2}
S^{(2)}_{FP}[\phi]=S^{(2)}[A]+{\bar C}^a\pa^{\mu}D^{ab}_{\mu}(A)C^b+B^a\chi^a(A,B),
\eeq
where
$\phi^{A}=\big(A^{a\mu},C^a,{\bar C}^a,B^a\big)$ are fields of full antisymplectic
space for Yang-Mills theory in the second order formulation.
The cases $\xi=0$ for (\ref{FP2}) and  for (\ref{FP1}) correspond
to singular gauges
being useful for theoretical treatments of quantum properties
while for practical quantum calculations the case $\xi\neq 0$
is more preferred.

The action $S_{FP}^{(1)}[\phi^{(1)}]$ is invariant,
\beq
\delta_BS_{FP}^{(1)}[\phi^{(1)}]=0,
\eeq
under the BRST transformations,\footnote{For more compact presentation we
use notation $\delta_B$ for $\delta_{BRST}$.}
\beq
\nonumber
&&\delta_B A^a_{\mu}=D^{ab}_{\mu}(A)C^b\lambda,\quad
\delta_B {\cal F}^a_{\mu\nu}=f^{abc}{\cal F}^b_{\mu\nu}C^c\lambda,\\
&&
\label{brst1}
\delta_B C^a=-\frac{1}{2}f^{abc}C^bC^c\lambda,\quad
\delta_B {\bar C}^a=B^a\lambda,\quad
\delta_B B^a=0,
\eeq
where $\lambda$ is constant Grassmann parameter.
In its turn the action $S^{(2)}_{FP}[\phi]$ is invariant,
\beq
\delta_B S^{(2)}_{FP}[\phi]=0,
\eeq
under standard BRST transformations \cite{brs1,t}
\beq
\label{brst2}
\delta_B A^a_{\mu}=D^{ab}_{\mu}(A)C^b\lambda,\quad
\delta_B C^a=-\frac{1}{2}f^{abc}C^bC^c\lambda,\quad
\delta_B {\bar C}^a=B^a\lambda,\quad
\delta_B B^a=0.
\eeq
The BRST transformations (\ref{brst1}) and (\ref{brst2}) play important role on quantum level
in deriving
the Ward identities for generating functionals of Green functions.

\section{Quantization}
\noindent
The generating functional of Green functions of fields $A^a_{\mu}$ and
${\cal F}^a_{\mu\nu}$ for the Yang-Mills theory in the first order form reads
\beq
\label{Z1}
Z^{(1)}[j,J]=\int D\phi^{(1)}
\exp\Big\{\frac{i}{\hbar}\big(S_{FP}^{(1)}[\phi^{(1)}] +jA+J{\cal F}\big)\Big\},
\eeq
where $j=\{j^a_{\mu}\}$ and $J=\{J^a_{\mu\nu}\}$ are external sources to fields
$A^a_{\mu}$ and ${\cal F}^a_{\mu\nu}$ respectively and the following abbreviations
\beq
jA=j^a_{\mu}A^{a\mu},\quad
J{\cal F}=J^a_{\mu\nu}{\cal F}^{a\mu\nu}
\eeq
are used.

Making use of change of integration variables in the form of the BRST transformations
(\ref{brst1}) and taking into account that the corresponding Jacobian is trivial due to
antisymmetry property of $f^{abc}$ we derive the relation
\beq
\label{wi1}
j^a_{\mu}\langle D^{ab\mu}(A)C^b\rangle^{(1)}_{j,J}+
J^a_{\mu\nu}f^{abc}\langle {\cal F}^{b\mu\nu}C^c\rangle^{(1)}_{j,J}=0,
\eeq
where the symbol $\langle (\cdots )\rangle^{(1)}_{j,J}$ means vacuum average of quantity
$(\cdots)$ in presence of external sources,
\beq
\label{vep1}
\langle (\cdots) \rangle^{(1)}_{j,J}=
\int D\phi^{(1)} (\cdots)\exp\Big\{\frac{i}{\hbar}
\big(S_{FP}^{(1)}[\phi^{(1)}] +jA+J{\cal F}\big)\Big\}.
\eeq
Switching off the sources $j,J$ in (\ref{vep1}) gives the corresponding Green functions of
Yang-Mills theory in the first order formulation. The relation (\ref{wi1}) is nothing but
the Ward identity formulated in the form without introduction of additional sources
to ghost and antighost fields and to generators of the BRST transformations. Any relations
existing among Green functions of Yang-Mills theory in the first order form
can be derived from the Ward identity by differentiating
with respect to sources $j^a_{\mu}$ and $J^a_{\mu\nu}$ and
then putting $j^a_{\mu}=0$ and $J^a_{\mu\nu}=0$.

The generating functional of Green functions of fields $A^a_{\mu}$
for Yang-Mills theory in the second order form is defined as
\beq
\label{Z2}
Z^{(2)}[j]=\int D\phi
\exp\Big\{\frac{i}{\hbar}\big(S_{FP}^{(2)}[\phi] +jA\big)\Big\}.
\eeq
The BRST symmetry of $S_{FP}^{(2)}[\phi]$ leads to the Ward identity
\beq
\label{wi2}
j^a_{\mu}\langle D^{ab\mu}(A)C^b\rangle^{(2)}_{j}=0,
\eeq
with the evident notation,
\beq
\langle (\cdots)\rangle^{(2)}_{j}=\int D\phi (\cdots)
\exp\Big\{\frac{i}{\hbar}\big(S_{FP}^{(2)}[\phi] +jA\big)\Big\}.
\eeq

Making use the change of integration variables in the functional integral (\ref{Z1})
\beq
{\cal F}_{\mu\nu }^{a}=E_{\mu\nu }^{a}+F_{\mu\nu }^{a}(A),
\eeq
we obtain
\beq
\label{Z1a}
Z^{(1)}[j,J]=\int D\phi
\exp\Big\{\frac{i}{\hbar}\big(S_{FP}^{(2)}[\phi] +jA\Big\}\Sigma[J,A],
\eeq
where the functional $\Sigma[J,A]$ is defined by the following functional integral
\beq
\label{Sigma}
\Sigma[A,J]=\int D E\exp\Big\{\frac{i}{\hbar}\Big(\frac{1}{4}
E_{\mu \nu }^ {a}E^{a\mu \nu }+
J^a_{\mu\nu}\big(E^{a\mu\nu}+F^{a\mu\nu}(A)\big)\Big)\Big\}.
\eeq
From (\ref{Sigma}) it follows that $\Sigma[A,J]$ does not depend on $A^a_{\mu}$
when $J^a_{\mu\nu}=0$, $\Sigma[A,0]=\Sigma$,
\beq
\label{Sigma0}
\Sigma=\int D E\exp\Big\{\frac{i}{4\hbar}
E_{\mu \nu }^ {a}E^{a\mu \nu }\Big\}={\rm const}.
\eeq
This fact plays important role in solving the quantum equivalence
for the Yang-Mills theory in the first and second order formulations.

\section{Quantum equivalence}
\noindent
Now we are in position to study quantum (non)equivalence. To do this we have
to compare vacuum functionals for the Yang-Mills theory formulated
in the first and second order
formalisms.  From (\ref{Z1a}) and  (\ref{Z2}) it follows that up to non-essential constant
$\Sigma$ (\ref{Sigma0})
\beq
Z^{(1)}[0,0] = Z^{(2)}[0],
\eeq
i.e. vacuum functionals coincide. It means the quantum equivalence of two schemes
of quantizations. Moreover there exists the relation
\beq
Z^{(2)}[j]=Z^{(1)}[j,0],
\eeq
which has been used in \cite{McFM-FB,McBFM-F,BFM-FMc} for calculations of Green functions
within the standard quantum
approach to the Yang-Mills theory with the help of the first order formulation.
The Ward identity (\ref{wi1}) reduces to (\ref{wi2}) when $J^a_{\mu\nu}=0$.

\section{Discussion}

\noindent
In the paper we have analyzed the Yang-Mills theory in the first order formulation.
This formulation operates with two set of fields $\{A^a_{\mu}\}$
and $\{{\cal F}^a_{\mu\nu}\}$ instead of $\{A^a_{\mu}\}$ in
the second order formalism. On classical level both formulations
are equivalent. Equivalence on quantum level needs in special study.
We have studied gauge symmetry of classical action in the first order formulation and
found two types of gauge transformations. One of them can be considered as
natural extension of gauge transformations of vector fields $A^a_{\mu}$ in the second order
formulation. The second type of gauge symmetry belongs to the class of trivial gauge
symmetry and doesn't correspond to additional degeneracy of initial action.
To construct full quantum action we have used the BV-formalism based on solution
to classical master equation and gauge fixing procedure \cite{BV,BV1}.  With the help of
full quantum action the generating functional of Green functions in the first order
formulation has been constructed and presented in the form of functional integral over
fields of full antisymplectic space. Comparing the vacuum functional in
the first order formulation with the vacuum functional in the second order form
the quantum equivalence of both quantization schemes has been proved.
A possibility to construct all Green functions within the second order formulations
with the help of the generating functional of Green functions of the first order approach
to the Yang-Mills theory used widely in \cite{BMc,BFMc1,McFM-FB,McBFM-F,BFM-FMc} was confirmed.

\section*{Acknowledgments}
\noindent
The author thanks I.L. Buchbinder and I.V. Tyutin for useful discussions and F.T. Brandt
with correspondence.
The work  is supported
by Ministry of  Education of Russian Federation,
project FEWF-2020-0003.

\end{document}